\documentclass[twocolumn,letter]{jpsj3}
\usepackage{bm}
\usepackage{color}
\usepackage{enumerate}

\title{Valley Spin Sum Rule for Dirac Fermions: Topological Argument}
\author{Jun \textsc{Goryo}
\thanks{E-mail address: jungoryo@iis.u-tokyo.ac.jp}}

\inst{Institute of Industrial Science, The University of Tokyo, 4-6-1 Komaba, Meguro, Tokyo 153-8505, Japan}

\abst{
We consider a two-dimensional lattice system with two sites in its unit cell. 
In such a system, the Bloch band spectrum can have some valley points, around which Dirac fermions appear 
as low-energy excitations. Each valley point has a valley spin $\pm1$. 
In the system, there are two topological  numbers counting vortices and merons in the Brillouin zone, respectively.  
These numbers are equivalent, and this fact leads to a sum rule that states that the total sum of the valley spins is absent even in a system 
without time-reversal and parity symmetries. 
We can see some similarity between the valley spin and chirality in the Nielsen-Ninomiya no-go theorem in odd-spatial dimensions.}

\kword{valley spin, sum rule, 
Chern Integer and vortex, Dirac fermion and meron, parity anomaly, graphene, zero-gap organic conductor, topological insulator, Nielsen-Ninomiya no-go theorem}

\begin{document}
\maketitle

The Bloch band spectrum in some lattice systems have valley points, around which Dirac fermions appear as low-energy excitations.  
The Dirac fermion has attracted much attention over the past few decades, since  
it is closely related to the quantized Hall effect\cite{Deser-Jackiw-Templeton,Niemi-Semenoff,Redlich,Ishikawa,Semenoff,HaldaneQHE,Oshikawa}, and also plays important roles in the argument for topological insulators\cite{Hasan-Kane}.  
The low-energy electronic features of graphene\cite{Gaim-Novoselov} and the zero-gap organic 
conductor $\alpha$-(BEDT-TTF)$_2$I$_3$\cite{alpha-bedt-ttf,Kobayashi} can be well described by gapless Dirac fermions.   

In this study, we consider a two-dimensional lattice system with two sites in its unit cell and discuss valley points. 
Each valley point has a valley spin $\pm1$,   
which is well-defined as long as the intervalley mixing can be neglected, i.e., in the long-wavelength limit. 
We focus on the relation among the valley spin and two kinds of topological numbers 
counting vortices\cite{TKNN,Kohmoto-85} and merons\cite{Volovik,HgTe-theory} in the Brillouin 
zone, respectively.
It has been shown generally that these numbers are equivalent on a two-torus.\cite{Hatsugai-Ryu} 
Using long-wavelength formalism, we show that this equivalence leads to the fact that the total sum of the valley spins is absent. 
This sum rule is obvious when the system preserves time-reversal or parity symmetry, since 
a valley spin flips under these symmetry transformations. In our argument,  
we do not introduce any assumptions on these symmetries. 
Namely, we can show the sum rule more generally. 
We also discuss some similarity to the Nielsen-Ninomiya 
no-go theorem for lattice fermions in odd-spatial dimensions\cite{Nielsen-Ninomiya}. 
We use the natural unit $\hbar=c=1$. 

We start from a tight-binding model on a two-dimensional lattice system with two sites in the unit cell: 
$H=\sum_{lm} t_{lm} c^\dagger_l c_m,$ where $c_l^\dagger$ ($c_l$) is the creation (annihilation) operator 
of an electron at the $l$-th site, and $t_{lm}$ is the hopping 
parameter between the $l$-th and $m$-th sites.  
In this system, we can introduce the sublattice spins  
$A$ and $B$ with the associated Pauli matrices 
${\bm \sigma}$, and the Bloch Hamiltonian can be written as 
$\mathcal{H}_{\bm k}=\epsilon_{{\bm k}}\sigma_0 + {\bm d}_{\bm k} \cdot {\bm \sigma}, 
\label{H_k}$ 
where ${\bm k}$ is the two-dimensional crystal momentum, $\sigma_0$ is the $2\times 2$ unit matrix, and  
${\bm d}_{\bm k}$ is a three-dimensional vector in the 
sublattice spin space. The real spin is omitted, i.e., we consider a system where the real spin is not activated.                                                                                                                                                                                                                                                                                                                                                                                                                                    The eigenvalue is 
$E_{\bm k}^{\pm}=\epsilon_{\bm k} \pm |{\bm d}_{\bm k}|$.   
We note that the condition for crossing between the upper  
and lower energy bands is $d_{1{\bm k}}=d_{2{\bm k}}=d_{3{\bm k}}=0$; and therefore, the most probable crossing is 
pointlike touching at some points in the two-dimensional Brillouin zone. More rigorous arguments are given in the two-band case\cite{Hatsugai-DP} 
and general case\cite{Asano-Hotta}. 
We use the following expression for the eigenstate: 
\begin{equation}
|u^{(\pm)}_{\bm k}\rangle
=\frac{1}{N_{\bm k}^\pm}\left(\begin{array}{c}
d_{3{\bm k}}\pm|{\bm d}_{\bm k}|
\\
d_{1\bm k} + i d_{2\bm k}
\end{array}
\right),
\label{eigenvector}
\end{equation}
where $N_{\bm k}^\pm$ is the normalization. 
We note the fact that we may choose the other expression $|u^{(\pm)}_{\bm k}\rangle \propto (-d_{1\bm k}+i d_{2\bm k}, d_{3\bm k}\mp|{\bm d}_{\bm k}|)^T$.

In such a model, there can be some valley points in the energy spectrum.   
Around the $i$-th valley point ${\bm k}_i$, the expansion of ${\bm d}_{\bm k}$ is given by 
${\bm d}_{{\bm k_i}+\delta {\bm k}}={\bm d}_{{\bm k}_i}+ \delta k_x \partial_{k_x}{\bm d}_{\bm k_i} 
+ \delta k_y \partial_{k_y}{\bm d}_{\bm k_i} 
+\mathcal{O}(\delta {\bm k}^2), $ 
where 
\begin{eqnarray}
{\bm d}_{{\bm k}_i} \cdot \partial_{k_x} {\bm d}_{{\bm k}_i}=
{\bm d}_{{\bm k}_i} \cdot \partial_{k_y} {\bm d}_{{\bm k}_i}=0  
\label{perp}
\end{eqnarray}
is satisfied [here, we introduced the notations $(\partial_{k_x},\partial_{k_y})=(\partial/\partial k_x,\partial/\partial k_y)$ and
$(\partial_{k_x} {\bm d}_{\bm k_i},\partial_{k_y}{\bm d}_{\bm k_i})=(\partial_{k_x} {\bm d}_{\bm k},\partial_{k_y} {\bm d}_{\bm k})|_{{\bm k}={\bm k_i}}$].
In principle, we can systematically find out all of 
the valley points when we obtain an explicit form of ${\bf d}_{\bm k}$ from the tight-binding Hamiltonian.

We show a procedure for obtaining 
the ``canonical form" of ${\bm d}_{\bm k}$, which leads to the Dirac equation directly. 
The argument presented here 
basically relies on that given by Oshikawa\cite{Oshikawa}. 
We consider the rotation $d^m_{\bm k} \rightarrow \sum_l d^l_{\bm k} R^{lm}_{\bm k}$, where $l,m=1,2,3$ denote the coordinates in the sublattice spin space. 
We see that the Hamiltonian is transformed to $\mathcal{H}_{\bm k} \rightarrow U_{\bm k} \mathcal{H}_{\bm k} U_{\bm k}^\dagger$, where $U_{\bm k}$ is 
a SU(2) matrix satisfying $\sum _m R^{lm}_{\bm k} \sigma^m=U_{\bm k} \sigma^l U_{\bm k}^\dagger$. Then, the state vector is transformed to $|u_{\bm k}^{(\pm)}\rangle \rightarrow U_{\bm k} |u_{\bm k}^{(\pm)}\rangle$. 
Using this rotation,  we can align ${\bm d}_{{\bm k}_i}$ along the ${\bm e}_3$ axes. 
It seems that we may direct ${\bm d}_{{\bm k}_i}$ parallel or antiparallel to ${\bm e}_3$; however, 
such an ambiguity can be removed as follows.  First, we let ${\bm d}_{{\bm k}_i} \rightarrow |{\bm d}_{{\bm k}_i}| {\bm e}_3$. 
We focus on the lower energy state (the discussion for the upper energy one is completely parallel), and 
($\mathcal{A}$) if $|u_{{\bm k}_i}^{(-)}\rangle \rightarrow (0, \exp[i \tan^{-1} (d_{2{\bm k}_i}/ d_{1{\bm k}_i})])^T$, 
it is consistent with the behavior of expression (\ref{eigenvector}) we have chosen. However, 
($\mathcal{B}$) if $|u_{{\bm k}_i}^{(-)}\rangle \rightarrow (0, 1)^T$, it is inconsistent. This behavior coincides with 
that of the other expressions for $|u_{{\bm k}_i}^{(-)}\rangle$. 
Thus, we flip ${\bm d}_{{\bm k}_i} \rightarrow -|{\bm d}_{{\bm k}_i}| {\bm e}_3$ 
[i.e., rotate $\pi$ around the ${\bm e}_2$ axes] 
and then $|u_{{\bm k}_i}^{(-)}\rangle \rightarrow (-1, 0)^T$, 
which is consistent with eq. (\ref{eigenvector}). 
After performing an appropriate transformation, we introduce the mass parameter 
$m_i=d_{3{\bm k}_i}$, which is positive in case ($\mathcal{A}$), and negative in case ($\mathcal{B}$). Obviously, $|m_i|$ gives the band gap at ${\bm k}_i$.

Now, the vectors $\partial_{k_x} {\bm d}_{{\bm k}_i}$ and $\partial_{k_y} {\bm d}_{{\bm k}_i}$
are in the plane perpendicular to ${\bm e}_3$ [see, eq. (\ref{perp}) ]. These two are not orthogonal, in general. 
Let us define the rank-2 tensor 
$A^{(i)}_{\mu\nu}=\partial_{k_\mu} {\bm d}_{{\bm k}_i}\cdot\partial_{k_\nu} {\bm d}_{{\bm k}_i}$, 
where $\mu,\nu=x,y$. This is a symmetric tensor. This tensor would be regular at a generic valley point. Thus, we can find out 
the principal axes $(k_{x^\prime}, k_{y^\prime})$ given by the rotation of the original axes $(k_x, k_y)$ [parity transformation $(k_x,k_y)\rightarrow (-k_x,k_y)$ is prohibited here, since it changes the valley spin 
defined below] 
and then $A^{(i)}_{\mu^\prime\nu^\prime}$ becomes diagonal, 
i.e., $\partial_{k_{x^\prime}} {\bm d}_{{\bm k}_i} \perp \partial_{k_{y^\prime}} {\bm d}_{{\bm k}_i}$. 
Moreover, we rotate the axes $(k_{x^\prime}, k_{y^\prime})$ so that $\partial_{k_{x^\prime}} {\bm d}_{{\bm k}_i} \parallel {\bm e}_1$ and $\partial_{k_{y^\prime}} {\bm d}_{{\bm k}_i}\parallel {\bm e}_2$. 
We define $v_{i \mu^\prime}=|\partial_{k_{\mu^\prime}} {\bm d}_{{\bm k}_i}|$ as the velocity. 
Then we arrive at the canonical form [hereafter, we omit the symbol $\prime$ for the principal axes]
\begin{eqnarray}
{\bm d}_{{\bm k}_i+\delta {\bm k}}=(\tau_i  v_{ix} \delta k_x, v_{iy}\delta k_y, m_i)+\mathcal{O}(\delta {\bm k}^2),    
\label{d}
\end{eqnarray}
where 
$\tau_i=\pm 1.$   
{\it This is the valley spin that denotes the relative sign between the coefficients of $\delta k_x$ and $\delta k_y$}. \cite{Gaim-Novoselov,alpha-bedt-ttf,Kobayashi} Here, we use the unitary transformation $U_{{\bm k}_i}=\sigma_3$ to set 
the coefficient of $\delta k_y$ positive. We should note that, in the above procedure, the parity transformations for ${\bm k}$ and the sublattice spin 
at a point on BZ should be prohibited, since we can change the sign of $\tau_i$ arbitrarily using these transformations.

The canonical form eq. (\ref{d}) leads to the Dirac equation, and the excitation near ${\bm k}_i$ is described by the Dirac fermion and becomes dominant when 
the band gap given by $2 |m_i|$ is sufficiently small compared with the bandwidth.  
The Dirac cone arises\cite{Gaim-Novoselov,alpha-bedt-ttf,Kobayashi} in the limit $|m_i|\rightarrow 0$. 
We also note that the vector eq. (\ref{d}) takes the configuration of a meron.\cite{Volovik,HgTe-theory}

In the limit of $\delta {\bm k} \rightarrow 0$, 
\begin{equation}
|u^{(-)}_{{\bm k}_i+\delta {\bm k}}\rangle
\rightarrow \left\{\begin{array}{cc}
\left(\begin{array}{c}
0
\\
\tau_i e^{i \tau_i \theta_{i \delta {\bm k}}}
\end{array} 
\right) & (m_i > 0), 
\\
\left(\begin{array}{c}
1 
\\
0
\end{array}
\right) & (m_i < 0), 
\end{array}
\right.
\label{eigenvector-}
\end{equation}
where $\theta_{i\delta {\bm k}}=\tan^{-1} (v_{iy} \delta k_y) / (v_{ix} \delta k_x)$. Namely, the lower-band state has a vorticity equal to 
$\tau_i=\pm1$ around ${\bm k}_i$ with $m_i>0$.  

We will see that the valley point specified by eq. (\ref{d}) gives elemental contributions to topological numbers, which will be discussed later. For 
simplicity, we assume that there is no point which gives higher contributions to the topological numbers. Such a point can be recognized 
as an overlap of valley points. 

Let us examine the sum rule 
\begin{eqnarray}
\sum_i \tau_i=0, 
\label{sum_rule}
\end{eqnarray}
where the summation is taken for all of the valley points in the band. This rule is obvious when the system preserves time-reversal and/or parity symmetry, since 
the valley spin is odd under time-reversal and parity transformations. Actually, this rule is satisfied 
in symmetric systems like graphene\cite{Gaim-Novoselov} and the zero-gap organic conductor 
$\alpha$-(BEDT-TTF)$_2$I$_3$\cite{alpha-bedt-ttf,Kobayashi}. 

We show the sum rule in a more general manner. We do not introduce any assumptions on time-reversal and parity symmetries. 
We consider two topological numbers for the lower energy band: 
\begin{eqnarray}
\mathcal{N}_{\rm vor}&=& \int_{BZ} \frac{d^2 k}{2 \pi i} {\bm \nabla}_{\bm k} \times {\bm A}_{\bm k},  
\label{vortex}\\
\mathcal{N}_{\rm mer}&=&\int_{BZ} \frac{d^2 k}{4\pi} \hat{\bm d}_{\bm k}\cdot \left(\partial_{k_x} \hat{\bm d}_{\bm k}\times \partial_{k_y} \hat{\bm d}_{\bm k}\right), 
\label{meron}
\end{eqnarray}
where 
${\bm A}_{\bm k}=\langle u_{\bm k}^{(-)} | {\bm \nabla}_{\bm k} | u_{\bm k}^{(-)} \rangle,$  
$\hat{\bm d}_{\bm k}={\bm d}_{\bm k}/|{\bm d}_{\bm k}|,$  
and $\int_{BZ} d^2 k$ denotes the integral over the {\it entire} Brillouin zone. 
These two numbers characterize the topological structure of the lower band.\cite{Comment_QHE} 
Generally, we can show that\cite{Hatsugai-Ryu} 
\begin{eqnarray}
\mathcal{N}_{\rm vor}=\mathcal{N}_{\rm mer}. 
\label{KUBO}
\end{eqnarray}

Let us estimate these numbers using the long-wavelength formations eqs. (\ref{d}) and (\ref{eigenvector-}).  
$\mathcal{N}_{\rm vor}$ is the Chern number that counts the total vorticity of the lower-band state\cite{TKNN,Kohmoto-85}, i.e.,  
\begin{eqnarray}
\int_{BZ} \frac{d^2 k}{2 \pi i} {\bm \nabla}_{\bm k} \times {\bm A}_{\bm k}&=&\sum_i \oint_{{\rm around}~{\bm k}_i} d {\bm k}\cdot  {\bm A}_{\bm k}
\label{N_vor}\\
&=& \sum_i \tau_i \theta(m_i)
=0,\pm 1,\pm 2,\cdot\cdot\cdot,
\nonumber
\end{eqnarray} 
where $\theta(x)$ is the step function. We can show it as follows: 
We may write ${\bm A}_{\bm k}={\bm A}_{\bm k}^{\rm vor}+\bar{{\bm A}}_{\bm k}$. 
The first part ${\bm A}_{\bm k}^{\rm vor}$ is defined to pick up the 
vortex singularity of the lower-band state eq. (\ref{eigenvector-}). Namely,  
\begin{eqnarray}
{\bm A}^{\rm vor}_{\bm k}=i {\bm \nabla}_{\bm k} \left\{\sum_i \tau_i  \theta(m_i) \tan^{-1} \frac{v_{iy}(k_y - k_{iy})}{v_{ix}(k_x-k_{ix})}\right\}, 
\end{eqnarray}
which gives the bottom line of eq. (\ref{N_vor}) \cite{Kohmoto-85}.
The remaining part $\bar{\bm A}_{\bm k}$ is, therefore, regular and does not contribute to $\mathcal{N}_{\rm vor}$,  
since the integral is defined on the entire Brillouin zone (two-torus).\cite{Kohmoto-85} Then, we obtain eq. (\ref{N_vor}).

On the other hand, $\mathcal{N}_{\rm mer}$ is related to merons in the Brillouin zone\cite{Volovik,HgTe-theory}. 
First, we vary all of the mass parameters $\{m_i | i=1,2...\}$ to be infinitesimal without closing the gaps, i.e., without sign changes. 
This deformation does not change $\mathcal{N}_{\rm mer}$ because of its topological nature. 
From the configuration of a meron shown in eq. (\ref{d}), we see that the integrand around ${\bm k}_i$ is 
\begin{eqnarray}
\hat{\bm d}_{\bm k}\cdot \left(\partial_{k_x} \hat{\bm d}_{\bm k}\times \partial_{k_y} \hat{\bm d}_{\bm k}\right)
=\frac{m_i \tau_i v_{ix} v_{iy}}
{(v_{ix}^2 \delta k_x^2 + v_{iy}^2 \delta k_y^2 + m_i^2)^{3/2}}, 
\label{ddddd}
\end{eqnarray} 
which is valid for a small $\delta {\bm k}$. To estimate its contribution to $\mathcal{N}_{\rm mer}$, we should introduce 
the appropriate momentum cutoff $\Lambda_{\delta k}$ around ${\bm k}_i$, which is comparable to the bandwidth on the energy scale. 
On the other hand, the integrand eq. (\ref{ddddd}) is localized strongly around ${\bm k}_i$, since its extension is characterized by an infinitesimally 
small $|m_i|$. 
Therefore, the contribution around ${\bm k}_i$ can be obtained from the integration of eq. (\ref{ddddd}) without a momentum cut off, which gives a half-quantized 
number, ${\rm sgn}(m_i) \tau_i/2$.  This is the so-called ``parity anomaly"\cite{HgTe-theory,Deser-Jackiw-Templeton,Niemi-Semenoff,Redlich,Ishikawa,Semenoff,HaldaneQHE,Oshikawa,Volovik}. Then, $\mathcal{N}_{\rm mer}$ is given by the sum of contributions from all valley points, i.e., 
\begin{eqnarray}
\int_{BZ} \frac{d^2 k}{4\pi} \hat{\bm d}_{\bm k}\cdot \left(\partial_{k_x} \hat{\bm d}_{\bm k}\times \partial_{k_y} \hat{\bm d}_{\bm k}\right)
= \sum_i \frac{1}{2}{\rm sgn}(m_i) \tau_i. 
\label{N_mer}
\end{eqnarray}
Eq. (\ref{N_mer}) has been verified numerically for arbitrary values of mass parameters 
using an explicit form of ${\bm d}_{\bm k}$ on the entire Brillouin zone in a certain tight-binding model.\cite{Imura}

We can see from eqs. (\ref{KUBO}), (\ref{N_vor}), and (\ref{N_mer}) that the number of valley points should be even.
We suppose that there are two valley points ${\bm k}={\bm k}_1, {\bm k}_2$. 
The extension to a system with $2N$ valley points ($N=2,3,4\cdot\cdot\cdot$) is straightforward. We emphasize that 
time-reversal and parity symmetries are not required here: 
We do not introduce any restrictions on the locations of valley points or on the values of the mass parameters. 
Below, we show that eqs. (\ref{N_vor}) and (\ref{N_mer}) give the same result and become consistent with eq. (\ref{KUBO}) 
when (I) the sum rule (\ref{sum_rule}) is satisfied, but do not when (II) the sum rule is not satisfied. 

We examine case (I) first. We can put $\tau_1=+1$ and $\tau_2=-1$ without losing generality. 
From eq. (\ref{N_vor}),  we obtain 
\begin{eqnarray}
\mathcal{N}_{\rm vor}=
\left\{\begin{array}{rl}
\tau_1+\tau_2=0 & m_1, m_2 >0, 
\\
\tau_1=+1 & m_1>0, m_2<0, 
\\
\tau_2=-1 & m_1<0, m_2>0, 
\\
0  & m_1, m_2<0.  
\end{array}
\right.
\label{vorticity}
\end{eqnarray}
On the other hand, from eq. (\ref{N_mer}), we obtain  
\begin{eqnarray}
\mathcal{N}_{\rm mer}&=&\frac{1}{2} \left\{\frac{m_1}{|m_1|}\tau_1 + \frac{m_2}{|m_2|}\tau_2\right\} 
\nonumber\\
&=&\left\{\begin{array}{ll}
0 & m_1, m_2 >0, 
\\
+1 & m_1>0, m_2<0, 
\\
-1 & m_1<0, m_2>0, 
\\
0  & m_1, m_2<0.  
\end{array}
\right.
\label{meron_charge}
\end{eqnarray}
Namely, $\mathcal{N}_{\rm vor}=\mathcal{N}_{\rm mer}$. 
In case (II), we immediately see that $\mathcal{N}_{\rm vor}\neq \mathcal{N}_{\rm mer}$, which is inconsistent with eq. (\ref{KUBO}).

To sum up, we have shown that two topological numbers shown by eqs. (\ref{vortex}) and (\ref{meron}) estimated in the long-wavelength 
formalism give a result consistent with the general relation eq. (\ref{KUBO}), when eq. (\ref{sum_rule}) is satisfied. 

\begin{figure}[h]
	\centering
	\includegraphics[width=9cm,clip]{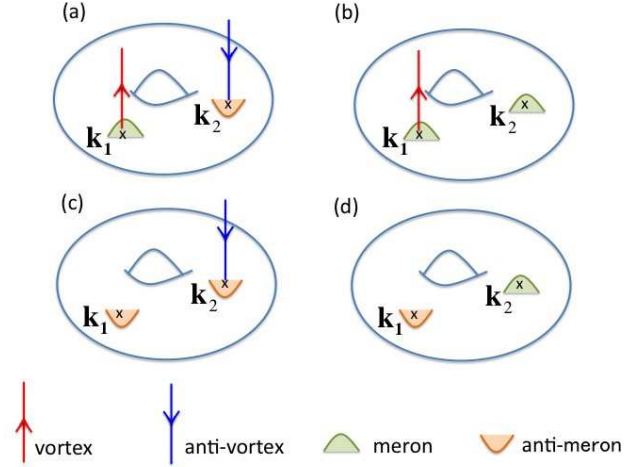} 
	\caption{(Color onlone) Configurations of vortices and merons on the Brillouin zone (depicted by the two-torus). 
	The valley spins are fixed at $\tau_1=+1$ and $\tau_2=-1$.  
	The signs of the masses are $m_1,m_2>0$ in (a), $m_1>0,m_2<0$ in (b), $m_1<0,m_2>0$ in (c), and $m_1,m_2<0$ in (d).}
	\label{vortex-meron}
\end{figure}

Let us discuss the configurations of vortices and merons, which are schematically shown 
in Fig. \ref{vortex-meron}. We note again that the valley spins are fixed at $\tau_1=+1$ and $\tau_2=-1$.
We note that a vortex with a vorticity equals to $\tau_i$ is located at ${\bm k}_i$ with $m_i>0$ [see eq. (\ref{eigenvector-})],  
on the other hand, a meron with a fractional charge ${\rm sgn}(m_i) \tau_i/2$ is located at each ${\bm k}_i$. 
Therefore, when (a) $m_1, m_2>0$, a vortex and a meron are located at ${\bm k}_1$, while 
an antivortex and an antimeron are located at ${\bm k}_2$. In the case that (b) $m_1>0, m_2<0$, a vortex is present 
at ${\bm k}_1$ but absent from ${\bm k}_2$, while a meron is located at ${\bm k}_1$ and ${\bm k}_2$. 
In the case that (c) $m_1<0, m_2>0$, an antivortex is present 
at ${\bm k}_2$ but absent from ${\bm k}_1$, while an antimeron is located at ${\bm k}_1$ and ${\bm k}_2$. 
When (d) $m_1, m_2<0$, there are no vortices in the entire Brillouin zone;  
however, a meron and an antimeron are located at ${\bm k}_2$ and ${\bm k}_1$, respectively.

The phases (b) and (c) in Fig. \ref{vortex-meron} have nonzero topological numbers and break parity and time-reversal symmetries.\cite{Comment_QHE}   
Such phases emerge if some interactions give $m_1$ and $m_2$ with opposite signs. 
Actually, this mechanism has been proposed by Haldane\cite{HaldaneQHE}, and its 
extension to the (time-reversal invariant) quantum spin Hall effect is given by Kane and Mele using a spin activating interaction\cite{KM1,KM2}. 
As long as the real spin is not activated, a system with parity and/or time-reversal symmetry is categorized 
into zero-topological number phase (a) or (d) in Fig. \ref{vortex-meron}. We see that graphene\cite{Gaim-Novoselov} and the organic 
conductor\cite{alpha-bedt-ttf,Kobayashi} are the gapless limits of phase (a) or (d). 

 The phase transition specified by the jump of the topological numbers 
occurs when $m_i$ changes its sign\cite{HgTe-theory,HaldaneQHE,Oshikawa}. 
For instance, let us see a transition from phase (d) to phase (b) in Fig. \ref{vortex-meron}, where the jumps of the topological numbers are 
$\Delta \mathcal{N}_{\rm vor}=\Delta \mathcal{N}_{\rm mer}=1$. We change $m_1$ continuously 
from a negative value to a positive value. 
At the transition point $m_1=0$, a Dirac cone appears at ${\bm k}_1$ in the energy spectrum\cite{HaldaneQHE}, 
and a vortex and a meron are created and an anti-meron is annihilated.  
The other Dirac-type spectrum at ${\bm k}_2$ with an opposite valley spin remains massive  
and hidden in the higher-energy part of the spectrum when its mass is comparable to the 
bandwidth.\cite{HaldaneQHE,Hatsugai-Kohmoto-Wu}

Let us discuss some relations to the Nielsen-Ninomiya no-go theorem on the lattice fermion doubling in 
odd-spatial dimensions\cite{HaldaneQHE,Hatsugai-DP,Nielsen-Ninomiya,Hatsugai-Kohmoto-Wu}. 
When the system possess parity symmetry, the sum rule eq. (\ref{sum_rule}) becomes obvious as we mentioned earlier, 
and the number of valley points should be even. Besides, the energy gap at a point is forbidden 
by the symmetry, since the mass term $m_i \sigma_z$ comes from the symmetry-breaking staggered potential.  
(Note that, at a valley point, the parity-invariant mass term $m_i \sigma_z s_z$ is allowed in graphene,\cite{KM1,KM2} but, we do not consider 
an interaction that activates the real spin) 
Therefore, we immediately see the fermion doubling.  
The role of the valley spin is similar to that of the chirality (the eigenvalue of $\gamma_5$ operator 
that can be defined in odd-dimensional space only) in the no-go theorem. A crucial difference is that the valley spin is still well-defined for a 
{\it massive} Dirac fermion as long as the intervalley scattering can be neglected; on the other hand, chirality is {\it not}. 
Our discussion would be related to the argument given in ref.\cite{Hatsugai-DP} in which a two-dimensional analog of ``chiral symmetry" is introduced artificially; however, the former appaers rather simpler.

The situation becomes somewhat indefinite in parity-symmetry-breaking systems. 
As we have shown, the sum rule (\ref{sum_rule}) is also satisfied in such systems. 
Owing to the sum rule (\ref{sum_rule}), if we found a valley point, there should be another point with an 
opposite valley spin. We assume the following: (A) Band gaps at these paired points $|m_1|$ and $|m_2|$ are degenerate, i.e., $|m_1 |=|m_2| \equiv m$. 
The presence of time-reversal symmetry is a sufficient condition for this degeneracy.\cite{note_sym2} 
(B) The gap amplitude $m$ is sufficiently smaller than the bandwidth. 
Under these assumptions, we find the statement:  
A massive Dirac fermion is always excited with its doubling partner that has 
{\it an opposite valley spin}, when the system has particle-hole symmetry (i.e., $\epsilon_{\bm k}=0$) and the Fermi level lies in the gap. 
Assumptions (A) and (B) would be, at least, approximately, satisfied when the symmetry-breaking 
perturbations are small.


The surface of a three-dimensional time-reversal invariant topological 
insulator\cite{Hasan-Kane,Fu-Kane,Qi-Hughes-Zhang} provides an exceptional case for the sum rule (\ref{sum_rule}). 
In such a system, the topological $\theta$-term exists in the bulk region as the hallmark of the $Z_2$ topological order, and has 
a surface term that coincides with the Chern-Simons term with 
{\it a half-quantized} Hall conductivity. 
This fact indicates that a single Dirac cone without a hidden partner in the higher-energy 
region exists at a time-reversal invariant point in the Brillouin zone for the surface state, since  
such a cone gives a half-quantized conductivity owing to 
parity anomaly\cite{Deser-Jackiw-Templeton,Niemi-Semenoff,Redlich,Ishikawa,Semenoff,HaldaneQHE,Oshikawa,Volovik,HgTe-theory} and 
matches with the presence of the Chern-Simons term. 
Thus, the sum rule fails. 
The topological connection between the bulk and boundary 
regions, the so-called ``bulk-boundary correspondence",\cite{Hasan-Kane,Fu-Kane,Qi-Hughes-Zhang} 
causes this unique situation. This is somewhat similar to the fact that 
the doubling partner mentioned in the no-go theorem\cite{Nielsen-Ninomiya} with opposite chirality  
is absent in the one-dimensional edge state of the quantized Hall effect 
where the bulk-boundary correspondence is also at work\cite{Wen,Hatsugai}. 

In summary, we have pointed out that the equivalence of two topological numbers [see eqs. (\ref{vortex}), (\ref{meron}), and (\ref{KUBO})] 
leads to the sum rule (\ref{sum_rule}) for the valley spin defined in the long-wavelength formalism. 
The sum rule is obvious when the system preserves time-reversal and/or parity symmetry, since a valley spin 
is odd under these symmetry transformations. In this study, the sum rule has been shown 
independently of the presence or absence of these symmetries. 
Basically, the valley spin at each valley point is determined by the detailed structure of the lattice tight-binding Hamiltonian. 
It seems interesting that a rigorous rule comes from 
a topological argument that is independent of the details of the Hamiltonian. 
We also emphasize that, to close the valley spin, we can see an analog of the fermion doubling theorem in odd-dimensional space
\cite{Nielsen-Ninomiya,Hatsugai-DP} in a rather simple manner.

The author thanks N. Hatano, D. S. Hirashima, A. Hotta, C. Hotta, T. Kawamoto, N. Maeda, T. Oka, A. Yamakage, and   
K.-I. Imura. The author also thanks 
the referee for suggesting the importance of 
Oshikawa's derivation\cite{Oshikawa} on the canonical form eq. (\ref{d}).
The author is financially supported by a Grant-in-Aid for Scientific Research 
from the Japan Society for the Promotion of Science under Grant 
No. 18540381, and also by the Core Research for Evolutional
Science and Technology (CREST) of the Japan Science and Technology Agency.


\begin{thebibliography}{99}
\bibitem{Comment_QHE} We comment on the fact that these numbers contribute to the Hall conductivity 
from the lower band when the band is fully filled at the zero temperature. \cite{TKNN,Kohmoto-85, Volovik,HgTe-theory}
\bibitem{note_sym2}  Time-reversal invariance requires $m_1=m_2$ when both of the valley points ${\bm k}_1$ and ${\bm k}_2$ are not 
on the time-reversal invariant points in the Brillouin zone (which is satisfied, in general), 
since ${\bm k}_1$ and ${\bm k}_2$ are replaced with each other by the symmetry transformation.
\end{thebibliography}
\end{document}